\documentclass[prl,twocolumn,superscriptaddress,floatfix,nobalancelastpage]{revtex4}

\usepackage{amsmath}
\usepackage{amssymb}
\usepackage{graphicx}
\usepackage{dcolumn}
\def \bisbse{(Bi$_{1-x}$Sb$_x$)$_2$Se$_3$}
\def \bise{Bi$_2$Se$_3$}
\def \bite{Bi$_2$Te$_3$}
\def \ten16{10$^{16}$cm$-3$}

\begin{document}

\title{Two-dimensional Dirac fermions in a topological insulator: transport in the quantum limit}

\author{J.G. Analytis}
\affiliation{Stanford Institute for Materials and Energy Sciences, SLAC National Accelerator Laboratory, 2575 Sand Hill
Road, Menlo Park, CA 94025, USA} 
\affiliation{Geballe Laboratory for Advanced Materials and Department of Applied
Physics, Stanford University, USA}

\author{R.D. McDonald}
\affiliation{Los Alamos National Laboratory, Los Alamos, NM 87545, USA}

\author{S. C. Riggs} \affiliation{National High Magnetic Field
  Laboratory and Department of Physics, Florida State University,
  Tallahassee FL 32310}

\author{J.-H. Chu}
\affiliation{Stanford Institute for Materials and Energy Sciences, SLAC National Accelerator Laboratory, 2575 Sand Hill
Road, Menlo Park, CA 94025, USA} 
\affiliation{Geballe Laboratory for Advanced Materials and Department of Applied
Physics, Stanford University, USA}

\author{G. S. Boebinger} \affiliation{National High Magnetic Field
  Laboratory and Department of Physics, Florida State University,
  Tallahassee FL 32310}

\author{I.R. Fisher}
\affiliation{Stanford Institute for Materials and Energy Sciences, SLAC National Accelerator Laboratory, 2575 Sand Hill
Road, Menlo Park, CA 94025, USA} 
\affiliation{Geballe Laboratory for Advanced Materials and Department of Applied
Physics, Stanford University, USA}

\begin{abstract}
Pulsed magnetic fields of up to 55T are used to investigate the
transport properties of the topological insulator \bise\, in the
extreme quantum limit. For samples with a bulk carrier density of n =
$2.9 \times 10^{16}$cm$^{-3}$, the lowest Landau level of the bulk 3D
Fermi surface is reached by a field of 4T. For fields well beyond this
limit, Shubnikov-de Haas oscillations arising from quantization of the
2D surface state are observed, with the $\nu = $1 Landau level
attained by a field of $\sim 35$T. These measurements reveal the
presence of additional oscillations which occur at fields
corresponding to simple rational fractions of the integer Landau
indices.
\end{abstract}


\maketitle

The recent prediction and discovery that \bise\, and \bite\, are
three-dimensional topological insulators (TI)
\cite{zhang_topological_2009,chen_experimental_2009,hsieh_tunable_2009,hsieh_observation_2009,alpichshev_stm_2010}
has sparked a flurry of investigations. In a magnetic field, their
relativistic dispersion causes the energy spectrum to be quantized so
that $E_{\nu}\propto \sqrt{B\nu}$, where $B$ is the magnetic field and
$\nu=1,2,3,..,$ is the energy level, known as the Landau level (LL).
The progression of energy levels $E_{\nu}$ has been recently observed
in scanning-tunneling microscopy (STM) on
\bise\cite{cheng_landau_2010,hanaguri_momentum-resolved_2010}. Other
experiments on bulk samples and nanoribbons have reported universal
conductance fluctuations attributed to the surface
state\cite{peng_aharonov-bohm_2010,checkelsky_giant_2009}. However,
all of the unambiguous measurements for the existence of the surface
state have been made by surface sensitive probes. In this letter, we
study samples of \bisbse\, in which the Fermi surface of the Dirac
fermions is small enough that pulsed fields of up to 55 T can access
their quantum limit - the limit in which only a few of the lowest LLs
are occupied. This is achieved by depleting the carrier density of the
bulk and thereby reducing the size of the Dirac Fermi
surface\cite{wray_observation_2009,analytis_bulk_2010}). We
demonstrate for the first time, a system in which not only the {\it
  transport} properties of Dirac fermions can be studied, but studied
in the 2D quantum limit where novel correlation effects are most
likely to arise\cite{ran_composite_2010,levin_fractional_2009}.

\begin{figure*}
\includegraphics[width =14.2cm ]{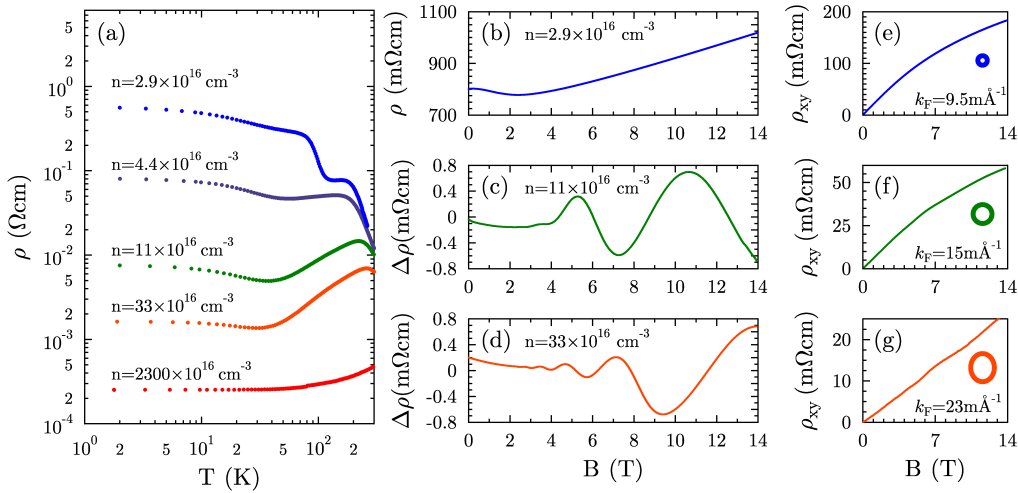}
\caption{(a) Progressively reducing the bulk carrier density in
  \bise\, results in a significantly increased resistivity. A
  reduction in the carrier density from n=$2300\times10^{16}$cm$^{-3}$
  to 33$\times10^{16}$cm$^{-3}$ was achieved by varying the growth
  conditions for a binary melt. Further reduction in the bulk carrier
  density down to n$=2.9\times10^{16}$ was achieved by isovalent
  substitution of Sb for Bi. (b-d) Representative magnetoresistance
  data at 2K for samples with indicated carrier densities, for field
  oriented along the trigonal c-axis. The progressive reduction of the
  Fermi surface volume is evident from the stretching of the SdH
  oscillations. (e-g) Carrier densities are extracted for each of
  these samples from the (low field) Hall effect. These are in very
  good agreement with those expected from the size of the Fermi
  surface (which shrinks monotonically as shown schematically in the
  lower right of each plot, with Fermi wavevector $k_F$
  indicated). For the lowest carrier density samples shown in panels
  (e) and (f), the Hall signal deviates from linearity as the field
  approaches the 3D (bulk) quantum limit. }
\label{byebyebulk} 
\end{figure*}

Generally speaking, transport measurements in \bise\, and \bite\, are
plagued by bulk conducting channels from either the conduction band or
by impurity bands introduced by foreign dopants
\cite{kasparova_n-type_2005,zhitinskaya_impurity_2004}. The materials
challenge is therefore finding a way to cleanly eliminate the bulk
conductivity so that the properties of the surface can be
observed. However, the carrier densities reported to date in studies
of TI remain only as
low\cite{chen_experimental_2009,hsieh_tunable_2009,hsieh_observation_2009,alpichshev_stm_2010}
as $\sim$10$^{18}$cm$^{-3}$. It has been known for some time that
substituting isovalent Sb for Bi can reduce the size of the Fermi
surface while retaining high mobilities
\cite{kulbachinskii_conduction-band_1999}.  By utilizing this idea and
an optimized growth technique \cite{analytis_bulk_2010} we have
reduced the bulk carrier density and corresponding Fermi surface
systematically, while increasing the bulk resistivity by several
orders of magnitude.

 Crystals with bulk carrier densities of 2300$\times 10^{16}$cm$^{-3}$
 and 33$\times 10^{16}$cm$^{-3}$ (red and orange curves on Figure 1
 (a)) were obtained by slow cooling a binary melt with different Bi:Se
 ratio. The principal origin of these relatively high carrier
 densities is Se deficiency and antisite defects. Samples with n$\leq
 11\times10^{16}$cm$^{-3}$ were obtained by slow cooling a ternary melt
 containing progressively more Sb. Although Sb is isovalent with Bi,
 Sb substitution apparently acts to control the defect density in the
 bulk crystals, reducing the bulk carrier density. Before measurements
 were performed, samples were cleaved on both sides with a scalpel blade
 in a dry atmosphere, mounted with contacts (using conductive silver
 epoxy) and pumped to 10$^{-4}$mbar within 20 minutes. The leads made
 contact along the sides of the crystal to maximize the contact with
 the bulk. All samples were measured using a standard 4-probe
 configuration. Magnetoresistance and Hall effect measurements were
 performed at the National High Magnetic Field Laboratory (Los Alamos)
 in the short pulse ($\sim$10ms rise time) 55T magnet. Sweeps at both
 negative and positive field polarities were measured for all
 temperatures and angular positions. The data were then `symmetrized'
 by combining the data from each field polarity to extract the
 components that are even in magnetic field (longitudinal
 magnetoresistance, $R_{xx}$) and odd in magnetic field (Hall
 resistance, $R_{xy}$).

In Figure \ref{byebyebulk} we illustrate the dependence of the
resistivity on temperature for various samples with different carrier
densities. The carrier density is determined by the (low field) Hall
effect and can be matched to the size of the Fermi surface measured by
quantum oscillations, which in resistivity are denoted Shubnikov-de
Haas oscillations (SdHO) (see Figure \ref{byebyebulk} (b) to
(d)). These oscillations are periodic in inverse field and their
period $\Upsilon$ can be related to the extremal cross sectional area
$A_k$ of the Fermi surface in momentum space via the Onsager relation
$1/\Upsilon=(\hbar/2\pi e)A_k$. When the field is rotated about the
crystal trigonal axis, the frequency of the SdHO changes according to
the morphology of the Fermi surface. In every one of our samples the
bulk Fermi surface is a closed three-dimensional ellipsoid
\cite{kohler_galvanomagnetic_1975,kohler_conduction_1973,analytis_bulk_2010}.
The effective mass $m^*\sim0.12m_e$ varies weakly with doping for
these carrier densities, as previously reported
\cite{kohler_galvanomagnetic_1975,kohler_conduction_1973}). For our
lowest carrier density sample, the pocket is expected to have an
orbitally averaged Fermi wavenumber of $k_F=0.0095\AA^{-1}$. With such
a small Fermi surface we are able to exceed the bulk 3D quantum limit
with moderate fields $\sim 4T$. The remainder of this study is
dedicated to the properties of the lowest carrier density samples at
high fields.

\begin{figure*}
\includegraphics[width = 14.2cm]{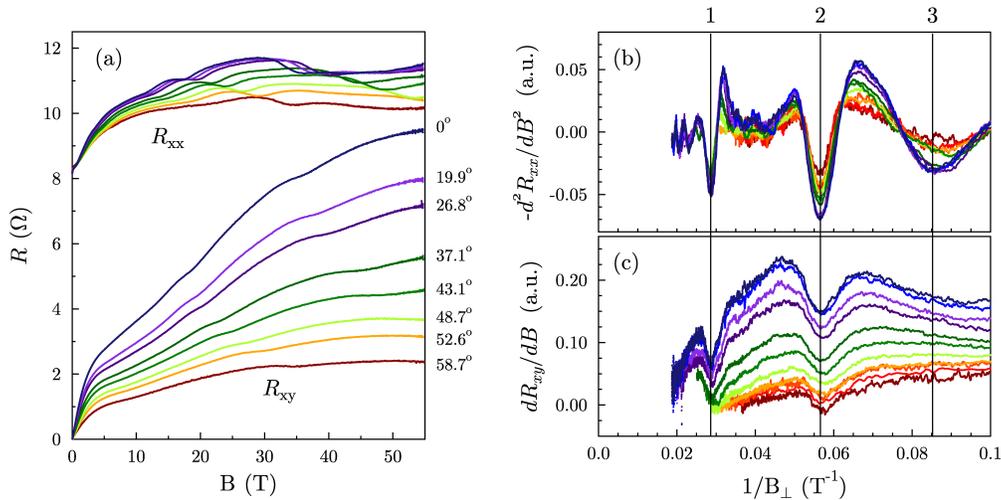}
\caption{\, (a) $R_{xx}$ and $R_{xy}$ data traces as a function of
  magnetic field for indicated angles for a single crystal with
  n$\sim3\times10^{16}$cm$^{-3}$.  The $R_{xx}$ traces are offset by
  $+$2.5 $\Omega$.  The deviation from linearity of the low-field
  $R_{xy}$ indicates the 3D quantum limit ($B\sim4$T). Beyond this
  limit, additional features in both $R_{xy}$ and $R_{xx}$ move
  smoothly up in field as the tilt angle $\theta$ is increased. In
  contrast, the 3D quantum limit is independent of tilt angle. (b, c)
  $-d^2R_{xx}/dB^2$ and $dR_{xy}/dB$ as a function of $1/B_{\perp}$,
  where $B_{\perp}=B$cos$\theta$ which aligns all features associated
  with the 2D surface state. The vertical lines are evenly spaced in
  1$/B_{\perp}$ and correspond to the filling of the first three
  energy levels $\nu$=1,2,3 of the 2D state.}
\label{hellosurface} 
\end{figure*}

In Figure \ref{hellosurface} (a) we illustrate the symmetrized data
for the longitudinal and transverse (Hall) resistances, $R_{xx}$ and
$R_{xy}$ respectively, taken at 1.5 K on a sample (of carrier density
n$\sim 3\times 10^{16}$cm$^{-3}$) of dimensions
$\sim$.63$\times$.36$\times$.15 mm$^3$. Strong features appear in the
$R_{xy}$ and $R_{xx}$ signal at similar fields
\cite{sarma_perspectives_1996}. To investigate the dimensionality of
the physics underlying these features, we rotate the crystal about the
field, defining $\theta=0$ to be when the field is perpendicular to
the surface (parallel to the trigonal c-axis). For the 2D surface
state of a topological insulator, quantum oscillatory phenomena depend
only on the perpendicular component of the field $B_{\perp}$, and
despite the peculiar Landau quantization $E_{\nu}$, the SdHO are still
periodic in $1/B$\cite{novoselov_room-temperature_2007}. In Figure
\ref{hellosurface} (b) and (c) we have plotted the $dR_{xy}/dB$ and
$-d^2R_{xx}/dB^2$ as a function of $B_{\perp}=B$cos$\theta$ - this
would be the usual procedure to determine whether minima in the
$R_{xx}$ (equivalent to minima in the negative second derivative) fall
on top of the Hall plateaus (appearing as minima in the first
derivative). Two pronounced dips in the Hall derivative are observed
to correspond with the same dips in the magnetoresistance at all
angles. The periodicity in 1$/B$ of the minima indicates they arise
from Landau quantization. In contrast to the angle dependence of the
bulk Fermi surface of higher carrier density samples
\cite{kohler_galvanomagnetic_1975,kohler_conduction_1973,analytis_bulk_2010}
these features scale with 1$/$cos$\theta$, providing unambiguous
evidence that the plateau-like features in the Hall and minima in the
SdHOs originate from a 2D metallic state.

\begin{figure}
\includegraphics[width = 9.2cm]{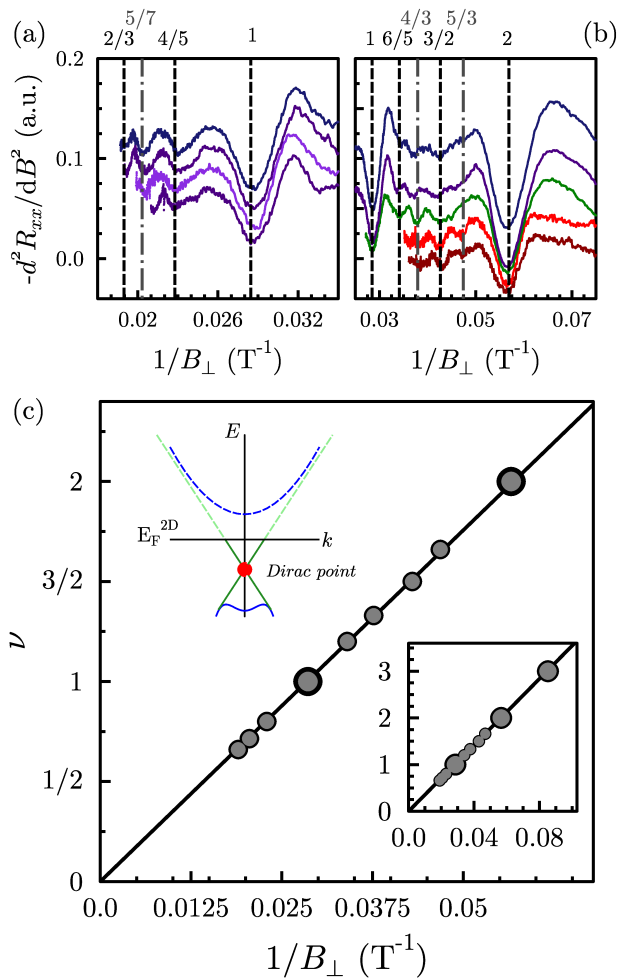}
\caption{ (a) $-d^2R_{xx}/dB^2$ as a function of $1/B_{\perp}$ in the
  field range for which $\nu < 1$ for $ \theta$ = 0$^\circ$,
  9.1$^\circ$, 19.9$^\circ$ and 26.8$^\circ$ (top to bottom),
  illustrating three additional features at fields corresponding to
  fractional landau indices.  (b) The same signal for the field range
  for which $1 < \nu < 2$ for angles $\theta$ = 0$^\circ$,
  19.9$^\circ$, 43.1$^\circ$, 55.7$^\circ$, 58.7$^\circ$, illustrating
  four further features. These additional oscillations scale with the
  the perpendicular component of the field to the surface just as in
  Figure \ref{hellosurface}, illustrating that they must also
  originate from the 2D surface state. Assigning indices for these
  features based on the integer indices gives values of
  $\nu=0.665\pm0.005,0.72\pm0.01,0.805\pm0.005$\, for $\nu<1$ and
  $1.19\pm0.02,1.32\pm0.02,1.51\pm0.03,1.65\pm0.03$\, for $1<\nu<2$,
  (counting from small to large 1$/B$). Vertical lines are drawn in
  panels (a) and (b) corresponding to the nearest fractional Landau
  filling. (c) LL index $\nu$ as a function of inverse field. Large
  symbols show integer levels $\nu=1,2,3$. Small symbols show
  fractional features observed in panels (a) and (b). The data are fit
  by straight line. The period of the integer states gives an estimate
  of the surface state $E_f$, which would cross 90 meV above the Dirac
  point corresponding to 110meV below the conduction
  band as shown in the schematic (upper left inset)
  \cite{analytis_bulk_2010}. }
\label{FQE} 
\end{figure}

 Examining Figure \ref{hellosurface} (b) the second and third minima
 occur at twice and three times the value in $1/B$ of the first, and
 it is natural to assign the indices $\nu=1,2,3$ as shown. We plot
 these indices as a function of 1$/B_{\perp}$ in Figure \ref{FQE} (c)
 which connect a straight line through the origin. The plateau-like
 features in $R_{xy}$ are reminiscent of the quantum Hall effect in 2D
 electron gases
 \cite{sarma_perspectives_1996,novoselov_room-temperature_2007}, but
 in the present case the quantization in integer multiples of
 $\sigma_0=e^2/h$ is not evident\cite{fu_topological_2007}. This can
 be attributed to the presence of a large parallel conductance channel
 from the bulk. Nevertheless, this data conclusively shows that we are
 able to get to the quantum limit {\it of the 2D state}.  The obvious
 candidate for the origin of the 2D states is the surface Dirac
 fermions.

\begin{figure}
\includegraphics[width = 7.2cm]{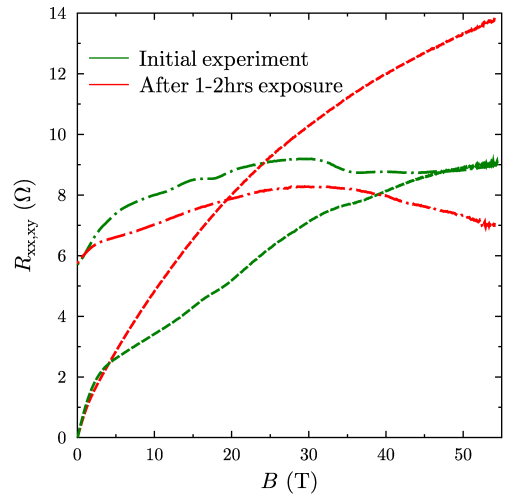}
\caption{ The $R_{xx}$ (dash-dot line) and $R_{xy}$ (dashed line)
  signal shown in green before exposure shows the oscillations related
  to the 2D state. The red curves show the signal after this sample is
  exposed to atmosphere for 1-2 hours, with greatly suppressed
  oscillatory signal in $R_{xx}$, and no apparent signature in
  $R_{xy}$.  }
\label{aging} 
\end{figure}

A crude method to confirm that the effect is surface related is to
briefly expose the sample to atmosphere. The atmosphere is a known
$n$-type dopant of the surface\cite{analytis_bulk_2010} so that bulk
carriers from the conduction band begin to contribute, smearing the
surface signal. After 1-2 hours exposure the oscillatory phenomena we
see in Figures \ref{hellosurface} and \ref{FQE} are almost completely
absent(see Figure \ref{aging}). We therefore conclude that the
observed signal does indeed originate from the surface state of the
TI. With this identification, we can determine the position of the
Fermi energy on the Dirac cone from the period $\Upsilon$ of the
oscillations. The Fermi wavevector $k_F$ of the surface state can be
calculated from the Onsager relation above, yielding
$k_F=0.031\AA^{-1}$. Using the Fermi velocity measured by
photoemission $v_F=4.2\times10^{5}$ms$^{-1}$ \cite{analytis_bulk_2010}, the
Fermi energy is estimated to be $\sim$90meV above the Dirac point, or
$\sim$110meV below the conduction band (see inset Figure \ref{FQE}
(c)).

It is important to distinguish this work from previous (unambiguous)
measurements of the Dirac fermion, which have been mostly done in zero
field\cite{chen_experimental_2009,hsieh_tunable_2009,hsieh_observation_2009,alpichshev_stm_2010}. The
high magnetic field lifts the degeneracy of the Dirac cone at the
Dirac point (defined in left inset Fig. \ref{FQE} (c)) via the spin
(Zeeman) coupling to the magnetic field, such that the quasiparticle
gains a mass\cite{shen_quantum_2009}. From the temperature dependence
of the quantum oscillations we can estimate a cyclotron mass
$m^*\propto dA_k/dE$ to be 0.11$m_e$ for the surface fermions. It
should be noted that even for a band with linear dispersion this value
is finite and related to the Fermi velocity $v_F$ and the band
filling\cite{novoselov_two-dimensional_2005}. In the present case
$v_F=4.2\times10^5$ms$^{-1}$ and $E_F=90$meV (see below) would give
$m^*=0.089m_e$. The difference with the measured mass provides a measure
of the band curvature of the split bands. This splitting fundamentally
changes the topology of the Dirac cone. In an unsplit system, like
graphene, the crossing at the Dirac point gives a Berry's phase factor
which results in a finite index intercept at
1/2\cite{mikitik_manifestation_1999,novoselov_two-dimensional_2005}.When
the Dirac cone is split (the fermions are
massive\cite{shen_quantum_2009}), this Berry's phase should not exist,
suggesting that the plot of $\nu$ vs $1/B$ extrapolates to
$(0,0)$\cite{mikitik_manifestation_1999}, as is presently observed
(see Figure \ref{FQE} (c)). Our data is therefore consistent with the
observation of this surface state with a split Dirac cone.

 In Figure \ref{FQE} (a) and (b) we plot $-d^2R_{xx}/dB^2$ vs
 $1/B_{\perp}$ in the range $0<\nu<1$ and $1<\nu<2$
 respectively. Minima in the signal emerge at each angle which line up
 when plotted as a function of $1/B_{\perp}$, and therefore also
 evidence 2D physics. These minima are not periodic in inverse field
 and cannot originate from another Fermi surface. They are also
 reproducible using different sweep rates of the pulsed magnetic field
 and thus cannot be artifacts of the pulsed magnet environment that
 are periodic in time. Intriguingly, these minima fall very near
 simple rational fractions of 2/3, 5/7 and 4/5 (in Fig. \ref{FQE} (a)
 as well as 6/5, 4/3, 3/2 and 5/3 (in Fig. \ref{FQE}(b)).

Aperiodic `oscillations' are typical of quantum interference phenomena
in mesoscopic systems, and we cannot rule out the possibility that
this is the origin of the additional non-periodic features seen in Fig
\ref{FQE} (a) and (b)\cite{checkelsky_giant_2009}. However, the
correspondence of these minima to simple fractions of the integer
Landau indices is striking, suggesting that they might be precursors
of the fractional QHE. The relatively low bulk mobility in the present
samples ($\sim$400cm$^2$/Vs) argues against such an interpretation,
but on the other hand, it is not clear whether the mobility of the
surface state is actually higher than that of the bulk. It is
instructive to note that the integer QHE in graphene was seen for
samples with mobilities around 40 times the present (bulk)
mobility\cite{novoselov_two-dimensional_2005}. The distinct steps
presently observed in $R_{xy}$ at integer Landau indices (reminiscent
of the integer QHE) argues for a higher surface mobility.

Except for the fraction $3/2$, the observed fractions indicated in
Fig.\ref{FQE} are where the classic {\it fractional} quantum Hall
effect (QHE) would be expected to be most robust. In semiconductors,
the strongest even denominator fraction occurs at 5/2, a state that is
both much weaker than odd-denominator
fractions\cite{willett_observation_1987} and is unusual in that it is
a spin-unpolarized FQH state \cite{eisenstein_collapse_1988}.  We note
that the data in Fig. \ref{FQE}(b) show the even denominator and odd
denominator fractions to be of similar strength.  If these features
are ascribed to the FQHE, this suggests that the strong spin-orbit
coupling in a TI may enable the Dirac quasiparticles to condense into
both symmetric and antisymmetric Laughlin
wavefunctions\cite{laughlin_anomalous_1983}.

By suppressing the bulk carrier density in \bisbse\, we have
demonstrated that the transport properties of the metallic surface
state of this topological insulator can be investigated in the quantum
limit. The application of high magnetic fields reveals quantum
oscillations arising from this new two-dimensional system, including
features at fields corresponding to {\it fractional} values of the
integer Landau indices.  As such, experimental access to the surface
state of \bisbse\, provides a new laboratory for studying topological
quantum matter and potentially correlations among Dirac fermions
\cite{ran_composite_2010,levin_fractional_2009}.

 The authors would like to thank Oscar Vafek, David Goldhaber-Gordon,
 Jimmy Williams, Shoucheng Zhang, Xiaoliang Qi, Chao-xing Liu, Joseph
 Maciejko, Yulin Chen, and Joel Moore for useful discussions.  The
 NHMFL is supported by NSF Division of Materials Research through
 DMR-0654118. RMcD acknowledges support from the U.S. DOE, Office of
 Basic Energy Sciences `Science in 100 T' program. This work is
 supported by the Department of Energy, Office of Basic Energy
 Sciences under contract DE-AC02-76SF00515.

\bibliographystyle{aps5etal}
%\bibliography{topoinsulators.bib}

\end{document}